\documentclass[twoside,english]{article}
\usepackage{lmodern}

\usepackage[T1]{fontenc}
\usepackage[latin9]{inputenc}
\usepackage{geometry}
\usepackage{float}
\usepackage{amsmath}
\usepackage{graphicx}
\usepackage{wasysym}
\usepackage{esint}

\makeatletter

\newcommand*\LyXZeroWidthSpace{\hspace{0pt}}
\let\SF@@footnote\footnote
\def\footnote{\ifx\protect\@typeset@protect
    \expandafter\SF@@footnote
  \else
    \expandafter\SF@gobble@opt
  \fi
}
\expandafter\def\csname SF@gobble@opt \endcsname{\@ifnextchar[%]
  \SF@gobble@twobracket
  \@gobble
}
\edef\SF@gobble@opt{\noexpand\protect
  \expandafter\noexpand\csname SF@gobble@opt \endcsname}
\def\SF@gobble@twobracket[#1]#2{}
\providecommand{\tabularnewline}{\\}

\@ifundefined{date}{}{\date{}}
\makeatother

\usepackage{babel}
\begin{document}
\title{\textbf{An Analysis of the Slayer Exciter Circuit}}
\maketitle
\begin{center}
Alen Kuriakose\footnote{alen.k@ahduni.edu.in}, Kharanshu Solanki\footnote{kharanshu.s@ahduni.edu.in},
Meblu Sanand Tom\footnote{meblu.t@ahduni.edu.in}
\par\end{center}

\begin{center}
\textit{School of Arts and Sciences, Ahmedabad University, Navrangpura,
Ahmedabad - 380009, India}
\par\end{center}

\begin{center}
Dated: November 27th, 2020
\par\end{center}
\begin{abstract}
\noindent In this paper, we try to create an effective mathematical
model for the well-known slayer exciter transformer circuit. We aim
to analyze various aspects of the slayer-exciter circuit, by using
physical and computational methods. We use a computer simulation for
data collection of various parameters pertaining to the circuit. Using
this data, we generate plots for various components and parameters.
We also derive an approximate equation to maximize the secondary output
voltage generated by the circuit. We also discuss a possible method
to construct such a circuit using low-cost materials.
\end{abstract}
\noindent \rule[0.5ex]{1\columnwidth}{0.2pt}

\tableofcontents{}

\listoffigures

\listoftables

\pagebreak{}

\section{Introduction}

Our aim throughout this paper, will be to analyze the slayer-exciter
circuit, as comprehensively as we possibly can. It can, at times,
be quite difficult to keep up with the structure of an academic paper.
Therefore, it is our deliberate intention to begin the paper with
an overview of the process we shall follow. \\

\noindent What does one require, in order to perform an experimental
analysis? Some may say we require computers, while others might say
we require apparatus. While it is true that we require an apparatus
to perform an experimental analysis, and that computational methods
can prove very useful in the contemporary scientific scenario, it
is also true (in certain scenarios) that we require a theory to test
in the first place. Of course, this isn't always the case, but for
this paper, it is. Therefore, we shall initiate the paper by a rather
brief discussion of the physical theories involved. These include
the likes of Lenz's law and the theory of transformers, among some
other very interesting theories.\\

\noindent The actual analysis lies in the computational model of the
circuit, for which, we will use the Falstad circuit simulation applet
{[}1{]}. This will allow us to collect data regarding various parameters
of the circuit, like the current, voltage and resistance. A large
enough data set will serve multiple purpose, i.e., of verifying already
existing theories, deriving a general equation to maximize the output
of the circuit, and most important of all, gaining understanding of
how various components of the circuit work together.

\section{Theory }

When an electron is in motion, i.e., if there is some current running
through some electrical component (like a wire), then an oscillating
electric field is produced {[}2-7{]}. This oscillating electric field,
in turn, generates an oscillating magnetic field. The planes of oscillation
of the electric and magnetic fields are orthogonal to each other (see
figure 1). This results in an electromagnetic wave which travels at
the speed of light c.

\begin{figure}[H]
\begin{centering}
\includegraphics[width=5cm,height=5cm,keepaspectratio]{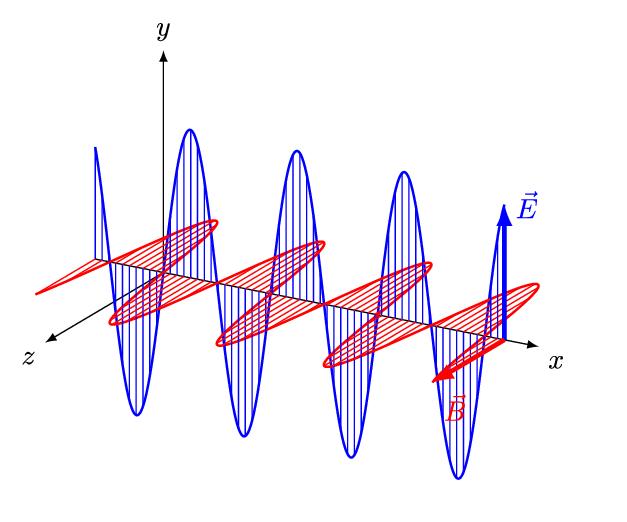}{\small{}\caption{An oscillating electric field produces an oscillating magnetic field,
and vice-versa, thereby resulting in an electromagnetic wave.}
}{\small\par}
\par\end{centering}
\end{figure}

\noindent Any wave traveling at speed $\upsilon$ can be mathematically
described as

\[
\nabla^{2}\psi-\frac{1}{\upsilon^{2}}\frac{\partial^{2}\psi}{\partial t^{2}}=0
\]

\noindent Therefore, for an electromagnetic wave, where the speed
is the speed of light $\mathit{c},$ we get,

\begin{align*}
\nabla^{2}\psi-\frac{1}{c^{2}}\frac{\partial^{2}\psi}{\partial t^{2}} & =0\\
\Rightarrow\left(\nabla^{2}-\frac{1}{c^{2}}\frac{\partial^{2}}{\partial t^{2}}\right)\psi & =0\\
\Rightarrow\Square^{2}\psi & =0
\end{align*}

\noindent Here, $\Square^{2}=\nabla^{2}-\frac{1}{c^{2}}\frac{\partial^{2}}{\partial t^{2}}$
is known as the d'Alembertian operator, which can be thought of as
a four-dimensional Laplacian operator {[}8-12{]}, with an additional
time-derivative component. 

\subsection{Faraday's Law of Electromagnetic Induction}

There are really two aspects to Faraday's law of electromagnetic induction.
These are formulated in terms of two laws, as follows:

\subsubsection{First Law}

Whenever a conductor is placed in a time-varying magnetic field, an
electromotive force (emf) is induced in it {[}2-7{]}. Then, if we
connect the conductor to a circuit and close it, a current is induced
in the circuit, called the induced current. The induced emf around
a closed loop $\mathit{L}$ is given as the line integral of the electric
field along the loop.
\begin{equation}
\varepsilon=\oint\overrightarrow{E}.d\overrightarrow{s}
\end{equation}
Here, $\varepsilon$ is the emf measured in volts (V), and $\overrightarrow{E}$
is the electric field generated.

\subsubsection{Second Law}

The electromotive force around a closed path is equal to the negative
of the time rate of change of magnetic flux through the surface which
the path encloses. For a loop of wire in a time-varying magnetic field,
the magnetic flux $\Phi_{B}$ is defined by a surface $\mathit{S}$,
whose boundary is given by a loop $\mathit{L}$. Since, the loop may
be moving, we can write $\mathit{S(t)}$for the surface. Then the
magnetic flux is the surface integral given {[}2-7{]} as

\begin{equation}
\Phi_{B}=\iint\overrightarrow{B(t)}.d\overrightarrow{A}
\end{equation}
Here, $\overrightarrow{B(t)}$ is the time-varying magnetic field
and $\overrightarrow{B(t)}.d\overrightarrow{A}$ is the dot product
showing the element of magnetic flux through a small element $\mathit{dA}$.
The magnetic flux through the loop will be proportional to the number
of magnetic field components passing perpendicular to the area enclosed
by the loop. Subsequently, the magnitude of the emf $\varepsilon$
induced in the circuit will be given according to Faraday\textquoteright s
law {[}2-7{]} of electromagnetic induction as
\begin{equation}
\varepsilon=\frac{d\Phi_{B}}{dt}
\end{equation}
If there are $\mathit{n}$ number of turns, or loops in the wire,
then we can just multiply the RHS of equation (3) by $\mathit{n,}$
and the equation will hold true.

\subsection{Lenz's Law}

The magnitude of the induced emf is given by the Faraday\textquoteright s
laws of electromagnetic induction. However, emf is not a scalar quantity.
It is a vector quantity, and therefore, there is a direction associated
with it, which is given by Lenz\textquoteright s law. Lenz\textquoteright s
law states that the current induced in a circuit due to induced emf
(due to changing magnetic field) is directed so as to oppose the change
in magnetic flux, and to exert a force which opposes the motion. Mathematically,
this adds ``direction'' to equations (3) and (4).

\begin{equation}
\varepsilon=-\frac{d\Phi_{B}}{dt}
\end{equation}

\noindent or in the case of $\mathit{n}$ loops,

\begin{equation}
\varepsilon=-n\frac{d\Phi_{B}}{dt}
\end{equation}

\noindent These equations basically imply that the back emf opposes
the changing current, which is the cause of the emf in the first place.

\subsection{Transformers and Inductance}

One of the most interesting features of Faraday\textquoteright s laws
is that a time-varying current in one coil can induce an emf in a
second coil placed close to it. Suppose that we take two coils, each
wound around separate bundles of iron sheets (these help to make stronger
magnetic fields), as shown in figure 2 below.

\begin{figure}[H]
\centering{}\includegraphics[width=5cm,height=5cm,keepaspectratio]{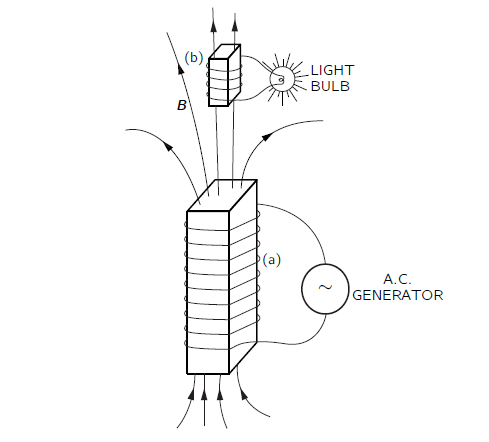}\caption{Two coils, wrapped around bundles of iron sheets, allow a generator
to light a bulb with no direct connection.}
\end{figure}

\noindent Now, when we connect one of the coils---coil (a)---to
an alternating-current generator. The continually changing current
produces a continuously varying magnetic field. This varying field
generates an alternating emf in the second coil---coil (b). This
emf can, for example, produce enough power to light an electric bulb.
The current in coil (b) can be larger or smaller than the current
in coil (a). The current in coil (b) depends on the emf induced in
it and on the resistance and inductance of the rest of its circuit.
The emf can be less than that of the generator if, say, there is little
flux change. Or the emf in coil (b) can be made much larger than that
in the generator by winding coil (b) with many turns, since in a given
magnetic field the flux through the coil is then greater. Such a combination
of two coils---usually with an arrangement of iron sheets to guide
the magnetic fields---is called a transformer. It can \textquotedblleft transform\textquotedblright{}
one emf (or voltage) to another. This effect is due to the mutual
inductance between the two coils. There are also induction effects
in a single coil. In figure 2, the varying current in coil (a) produces
a varying magnetic field inside itself and the flux of this field
is continually changing, so there is a self-induced emf in coil (a).
This effect is called \textit{self-inductance}. \\

\noindent The self-inductance $\mathcal{L}$ of a loop of wire carrying
a time-varying current $\mathit{I(t)}$, is given {[}2-7{]} as

\begin{equation}
\varepsilon=-\mathcal{L}\frac{dI(t)}{dt}
\end{equation}

\noindent Here, $\varepsilon$ is the electromotive force induced
in the loop. Now, for our purpose, we are more interested in the aspect
of mutual inductance, because that is one of the main principles of
a transformer. Figure 6 shows an arrangement of two coils, which demonstrates
the basic effects responsible for the operation of a transformer. 

\begin{figure}[H]
\centering{}\includegraphics[width=6cm,height=4cm,keepaspectratio]{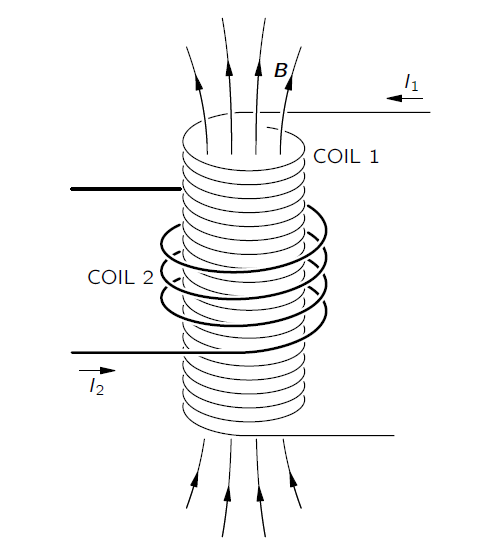}\caption{A current in coil 1 produces a magnetic field through coil 2.}
\end{figure}

\noindent Coil 1 consists of a conducting wire wound in the form of
a long solenoid. Around this coil---and insulated from it---is wound
coil 2, consisting of a few turns of wire. If now a current is passed
through coil 1, we know that a magnetic field will appear inside it.
This magnetic field also passes through coil 2. As the current in
coil 1 is varied, the magnetic flux will also vary, and there will
be an induced emf in coil 2. Suppose the current through coil 1 is
$\mathit{I_{1}}$, then the emf through coil 1 is given {[}2-7{]}
as

\begin{equation}
\varepsilon_{1}=-\mathcal{M}\frac{dI_{2}}{dt}
\end{equation}

\noindent Here, $\varepsilon_{1}$ is the emf induced in coil 1, $\mathit{I_{2}}$
is the current through coil 2, and $\mathcal{M}$ is the mutual inductance
of the two coils.

\subsection{Output Voltage of a Transformer}

Suppose we apply a certain amount of output voltage to the primary
coil of the transformer (figure 7 shows a basic step-down transformer
setup). This means an emf will be generated in the secondary coil
with due to mutual induction.

\begin{figure}[H]

\begin{centering}
\includegraphics[width=5cm,height=3cm,keepaspectratio]{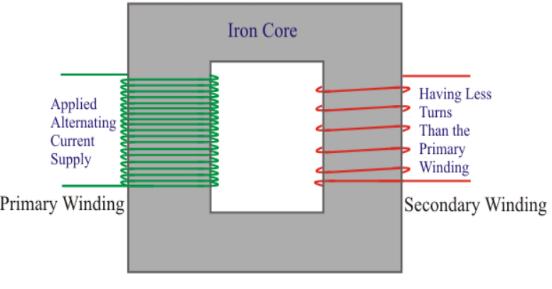}\caption{A basic step-down transformer.}
\par\end{centering}
\end{figure}

\noindent Now, for a step-down transformer, the number of turns of
secondary coil is less than the number of turns in the primary coil
($\mathit{n_{1}<n_{2}}$). The voltages across the primary and secondary
coils ($\mathit{V_{1}}$ and $\mathit{V_{2}}$ respectively) are related
to the number of turns ($\mathit{n_{1}}$ and $\mathit{n_{2}}$ respectively)
and the currents through them ($\mathit{I_{1}}$ and $\mathit{I_{2}}$
respectively) as,

\begin{equation}
\frac{V_{1}}{V_{2}}=\frac{n_{1}}{n_{2}}=\frac{I_{2}}{I_{1}}
\end{equation}

\noindent We will later verify equation (8), by analyzing the data
obtained from the computational simulation of the circuit. 

\subsection{Resonant Transformers}

There are various kinds of transformers that one can find. One of
the most efficient transformers is the resonant transformer. A resonant
transformer works on the principle of a resonance in the circuit.
\\

\noindent In simple terms resonance is basically the rapid pulsating
(switching on and off) of the current (or voltage) in the circuit.
But how does resonance of the circuit make the transformer more efficient?
Well, the resonance helps to maximise the voltage output of the transformer.
How? We will see that in a while. First, let\textquoteright s see
how resonance functions in AC and DC circuits. \\

\noindent Resonance is generally associated with an AC circuit (i.e.,
a circuit having an alternating voltage source). But the major problem
with using an AC source is that the resonance frequency of the source
(the frequency at which the AC source switches the circuit on and
off) is not the same as the resulting resonance frequency of the transformer
(the frequency at which the transformer coils pulsate). In order to
gain the maximum output voltage, these two resonant frequencies need
to be made equal in a process known as tuning of a transformer. This
tuning of a circuit needs to be done manually by adding some form
of external resonance generator to the circuit or by adding some extra
load to the transformer via a capacitor (see figure 8 below). 

\begin{figure}[H]
\begin{centering}
\includegraphics[width=6cm,height=3cm]{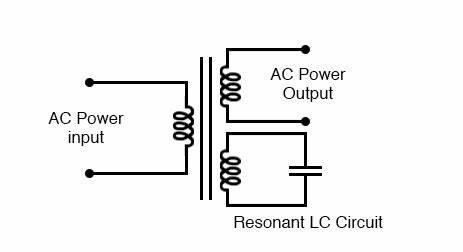}\caption{A capacitance is added to increase the load of the transformer in
order to match the frequency of the transformer coils to that of the
AC supply.}
\par\end{centering}
\end{figure}

\noindent We will discuss why the tuning of the transformer is important
later in this section. However, in the case of a DC circuit (like
the slayer exciter circuit), the resonance can be generated using
a transistor (which acts as a switch to rapidly turn on and off the
circuit). In this case, the transformer ends are connected to the
transistor and the resonant frequency of the transformer coils is
equal to the resonant frequency of the transformer coils. In other
words, the transformer is already auto-tuned! Hence, there is no need
for the extra hassle to manually tune the circuit. This is indeed
what makes a slayer exciter circuit unique in its working and distinguishes
it from other types of Tesla coils. But why do we need to tune the
transformer in the first place? Let\textquoteright s look at a famous
analogy to understand this. Suppose we have a simple pendulum (a bob
suspended by a string; see figure 9 below).

\begin{figure}[H]
\begin{centering}
\includegraphics[width=3cm,height=5cm,keepaspectratio]{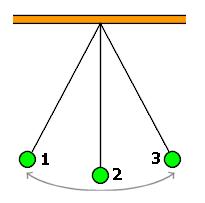}\caption{A simple pendulum. The bob (green) will reach maximum height only
if we keep applying forces at points 1 or 3.}
\par\end{centering}
\end{figure}

\noindent Our problem here is that we want to create the maximum possible
output, i.e., we want the pendulum bob to reach the maximum height
possible. Now imagine the pendulum was already oscillating at a certain
frequency (but the bob was not reaching the maximum possible height).
In order to maximise the height of the bob, we need to apply a force
to it. If we think about it deep enough, we will realize that we can\textquoteright t
apply the force at any random time. To generate the maximum height,
the force needs to be applied when the bob reaches the extreme points
in its trajectory (points 1 or 3 in figure 9). That\textquoteright s
the only way the bob can efficiently reach the maximum height. If
we apply a force when it is at the equilibrium point (coming towards
us or going away from us; point 2 in figure 9), the force applied
and the oscillating frequency of the spring won\textquoteright t be
in sync, and the result will not be maximised. In this analogy, the
oscillating frequency of the pendulum is analogous to the resonant
frequency of the AC source (in case of an AC circuit) or the resonant
frequency of the transistor (in case of a DC circuit); and the force
we apply to the pendulum is analogous to the resonant frequency of
the transformer coils.\LyXZeroWidthSpace\\

\noindent One can at least gain an intuition as to why the resonant
frequencies need to be equal (i.e., the transformer needs to be tuned)
in order to generate the maximum possible output voltage, from the
analogy discussed above. This discussion entails all the necessary
theory required to move on and perform an experimental analysis of
it. 

\pagebreak{}

\section{Making a Physical Model of the SEC\protect\footnote{SEC will be used as a substitute for the Slayer-Exciter Circuit.}}

While the functioning of a slayer exciter is quite simple to understand,
to actually make a physical model requires a significant amount of
cautiousness and patience. The following is the process we followed
to create the entire slayer exciter physical model\footnote{This method is probably the most preliminary one (or a low-cost one),
and can be vastly improved upon, using better instrumentation.}:
\begin{itemize}
\item Gather basic circuit components like n-channel MOSFETs, LEDs, zener
diodes, resistors, copper wires, and 9V batteries.
\item Use a plywood board to establish the base of the set-up.
\item Take a PVC pipe and wind a single strand copper wire around it (approx.
400 windings). This makes the secondary coil of our transformer. This
requires a lot of patience, because the winding needs to be perfect
in order to make the circuit function as expected.
\item Take a thicker copper wire and wind it around the secondary coil (approx.
3-5 windings). Do this as shown in figure 3.
\item Connect all the components together to form the circuit shown in figure
7 below (we will explain how these connections are made and how the
circuit works later in the report). Use a breadboard if necessary. 
\end{itemize}
\begin{figure}[H]
\centering{}\includegraphics[width=6cm,height=4cm,keepaspectratio]{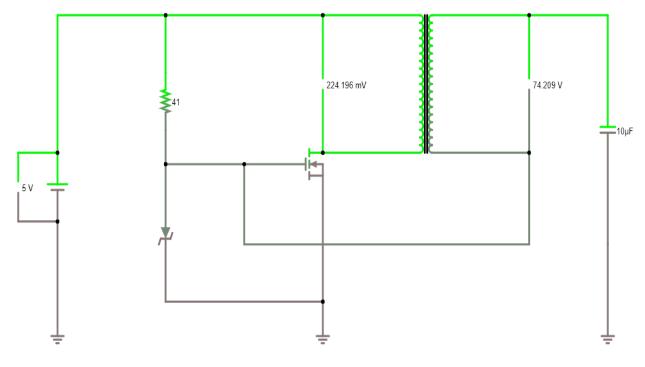}\caption{Basic schematic of the slayer exciter.}
\end{figure}

\begin{itemize}
\item Switch on the battery source.
\item Bring a small piece of conducting material close to the transformer.
Corona discharges may be observed. Corona discharges are the violet
sparks one can see due to the polarization of air surrounding the
high voltage transformer. The violet colour is due to the ionization
of nitrogen molecules in the air. So, if we bring a conducting material
close to the coil, we see a violet spark (corona discharge) between
the coil and the conducting material. This step demonstrates the working
of our physical slayer exciter circuit.
\end{itemize}

\section{Simulating the SEC}

Making the physical model had its own limitations. There was no way
we could verify the equation of a transformer (equation 8) using the
low-cost physical model, because the measurements would have had a
lot of uncertainty, and the results might have been inaccurate. Therefore,
we decided to verify the equations graphically using data from an
online simulation called Falstad Circuit Applet {[}1{]}. This applet
is a simulation software based on JavaScript and is very beneficial
to analyze any kind of circuit.

\begin{figure}[H]
\begin{centering}
\includegraphics[width=6cm,height=3cm,keepaspectratio]{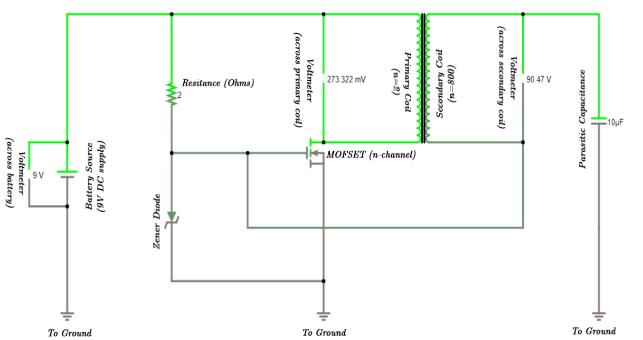}\caption{A simulated version of the slayer exciter circuit.}
\par\end{centering}
\end{figure}

\noindent Figure 8 above, shows the final simulated version of the
slayer exciter. Let us understand the flow of current in this circuit,
with the primary aim of understanding the roles of the different components
in the circuit. In figure 9 below, when the battery (B) is turned
on, the resistor R drives the current to the base of the transformer
(A). Consequently, A turns on and drives the current into the primary
coil (PC) of the transformer. 

\begin{figure}[H]
\centering{}\includegraphics[width=6cm,height=4cm,keepaspectratio]{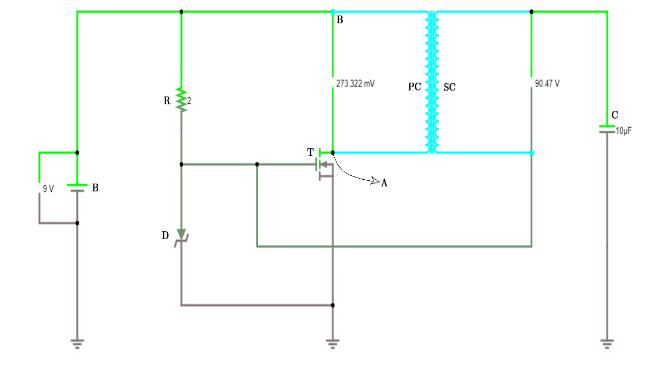}\caption{Circuit schematic for the slayer exciter.}
\end{figure}

\noindent The current is limited by the base current value. This creates
a magnetic field, which induces an emf, and therefore a current in
the secondary coil (SC) of the transformer. The secondary voltage
will tend to grow larger. But the tiny parasitic capacitance on the
output resists this change (although very small) against the rise
of the output end; and so, in return, the voltage on the other end
of the transformer goes down, pulling the base of the transistor low.
The role of the diode (D) is very important to the circuit. Since
we are using a DC input source, the diode prevents the base voltage
to fall more than a minimum specific value below the ground. This
in turn, step\textquoteright s up the output voltage of the secondary
coil (SC). As a result, the transistor (T) turns off and the magnetic
field starts to reduce. The base voltage increases and A turns on
again and the whole cycle keeps repeating (until the input is cut
off).\\

\noindent The first few questions that come in the mind after looking
at the schematics is that why have we used a MOSFET {[}13{]} instead
of a BJT (or for that matter, why have we used a n-p-n transistor
rather than a p-n-p transistor?); or why have we not used an LED instead
of a Zener diode? Well, the reason we used a MOSFET over a BJT, is
because MOSFETs are suitable for high voltage circuits like the slayer
exciter, because it has a very low drain-source. We saw how the current
switching (or resonance) of a circuit helps to maximise the output
voltage. The faster the switching process, the higher will be the
output voltage. Now, the current in a transistor is due to the flow
of majority charge carriers (electrons and holes). Electrons generally
travel faster than holes; and since the majority charge carriers for
n-p-n transistors are electrons, as opposed to p-n-p transistors (which
have holes as the majority charge carriers), we have preferred to
use a n-p-n transistor over a p-n-p transistor. \\

\noindent Coming to the diodes, the reason we did not use a simple
LED is because the input voltage is quite high (almost 14V), and an
LED will fuse out or burn out with a voltage supply as low as 4V!
In contrast, a Zener diode has a much higher input resistance and
can resist higher amounts of input voltage, which is the reason why
we used a Zener diode.\\

\noindent \textbf{Note:} We have provided a link to the simulation
text file in the references {[}14{]}. Instructions regarding how to
initiate the simulation are given in the text file itself. 

\section{Analyzing the SEC Simulation Data}

\subsection{Why Perform Analysis?}

Any kind of scientific report would be incomplete if it doesn\textquoteright t
provide some tangible results. The physical model of the slayer exciter
only serves the purpose of demonstrating its working. But that does
not help with any kind of numerical analysis of the circuit. What
if we want to know how to maximise the voltage output of the circuit?
There certainly isn\textquoteright t any way of doing that by just
looking at the physical model or by looking at its applications. Of
course, one could get voltmeters and ammeters and all kinds of output
meters to get some kind of data. But what if one doesn\textquoteright t
have all these resources? Well, for that very reason, we sought out
for some technique using which we could perform numerical analysis
without need of any kind of voltmeters or ammeters or stop-clocks.
The best way of doing so was by using the computer simulation we designed
in the previous section. We collected the data from the simulation,
embedded it into MATLAB, and generated graphical outputs for better
visualization.

\subsection{What Kind of Analysis?}

To serve the purpose of verifying theory, we have decided to plot
a graph of how the secondary output voltage varies with time. We will
also verify whether the ratio of primary and secondary voltages is
equal to the turn ratio of the coil. Barring these, our primary goal
is to find an equation that will tell us how to maximise the output
voltage by adjusting different parameters in the circuit like resistance,
input voltage, inductance, turn ratio and the parasitic capacitance.
Lets first see how the output voltage will vary with increasing time.

\subsection{Evolution of Output Voltage with Time}

We saw in previous sections that the transistor and the transformer
have resonant frequencies, which results in the rapid pulsating of
current in the circuit. This means that the current, and therefore
the voltage change directions as time passes. If we start at time
$\mathit{t=0s}$, and observe the change in voltage, we must expect
that the voltage traces out a sine curve. Table 1 shows the data for
secondary output voltage ($\mathit{V_{S}}$; in $\mathrm{V}$) w.r.t.
time ($\mathit{t}$; in $\mathit{s}$), and table 2 shows the values
of other parameters, for which the readings were taken.

\begin{table}[H]
\noindent \begin{centering}
\begin{tabular}{|c|c|c|c|c|c|c|c|}
\hline 
\textbf{Time ($\mathit{s}$)} & \textbf{Voltage (V)} & \textbf{Time ($\mathit{s}$)} & \textbf{Voltage (V)} & \textbf{Time ($\mathit{s}$)} & \textbf{Voltage (V)} & \textbf{Time ($\mathit{s}$)} & \textbf{Voltage (V)}\tabularnewline
\hline 
\hline 
0 1.004 & 0  & 12.002  & 155.175  & 24.045  & -36.5  & 36.041  & -146.464 \tabularnewline
\hline 
1.004 & 16.395 & 13.006 & 151.631 & 25.253 & -59.69  & 37.013  & -137.803 \tabularnewline
\hline 
2.042 & 38.879 & 14.003  & 145.094 & 26.005 & -74.618  & 38.005  & -126.28 \tabularnewline
\hline 
3.015 & 59.205 & 15.009  & 135.584 & 27.019 & -93.41 & 39.027 & -111.827 \tabularnewline
\hline 
4 & 78.643 & 16.039  & 122.997 & 28.02  & -110.061 & 40.056  & -94.018 \tabularnewline
\hline 
5.012 & 97.023  & 17.016 & 108.662 & 29  & -124.242 & 41.032  & -77.014 \tabularnewline
\hline 
6.089 & 114.401 & 18.058 & 91.077  & 30.014  & -136.401  & 42.017  & -57.465 \tabularnewline
\hline 
7.007  & 127.128 & 19.011  & 73.264 & 31  & -145.549  & 43.008 & -36.656 \tabularnewline
\hline 
8.001 & 138.496 & 20.056  & 52.234  & 32.007  & -151.962  & 44.022  & -14.65 \tabularnewline
\hline 
9.02  & 147.31 & 21.009  & 32.025  & 33.15  & -155.517  & 45.096  & 8.981 \tabularnewline
\hline 
10.01 & 152.948 & 22.004 & 10.395 & 34.011  & -155.48  & - & -\tabularnewline
\hline 
11.005  & 155.602 & 23  & -11.371  & 35.027 & -152.557  & - & -\tabularnewline
\hline 
\end{tabular}\caption{Data for $\mathit{V_{S}}$ vs $\mathit{t}$}
\par\end{centering}
\end{table}

\begin{table}[H]
\noindent \centering{}%
\begin{tabular}{|c|c|}
\hline 
\textbf{Parameter} & \textbf{Value}\tabularnewline
\hline 
\hline 
Resistance  & 22$\Omega$\tabularnewline
\hline 
Capacitance & 10$\mu F$\tabularnewline
\hline 
Primary Inductance & 5H\tabularnewline
\hline 
Input Voltage & 10V\tabularnewline
\hline 
Turns Ratio & 1000\tabularnewline
\hline 
\end{tabular}\caption{Values of constant parameters.}
\end{table}

\noindent One can easily notice, by looking at the data sets, that
the time intervals are not uniform. That is owing to one of the drawbacks
of the simulation, that the readings need to be taken manually by
pausing the simulation every now and then. Figure 10 below shows the
plot of secondary voltage w.r.t. time. As we expected, the nature
of the plot is that of a sine curve. Time (s) is plotted on the x-axis
and the secondary voltage (V) is plotted on the y-axis.

\begin{figure}[H]
\begin{centering}
\includegraphics[width=10cm,height=8cm,keepaspectratio]{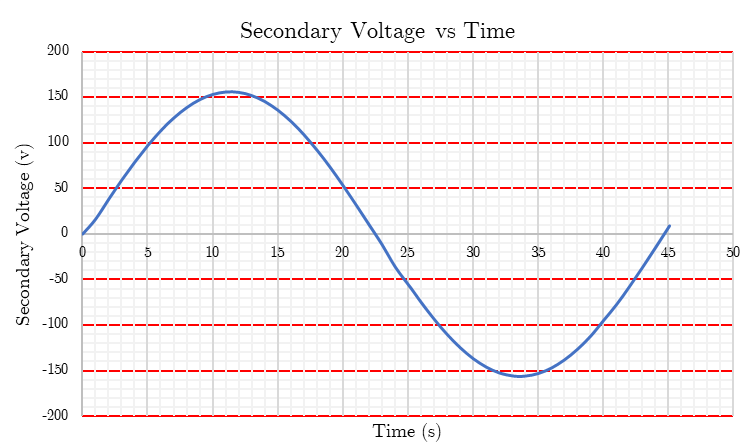}\caption{Plot of secondary output voltage versus time elapsed. The nature of
the graph is same as that of a sine curve.}
\par\end{centering}
\end{figure}

\noindent An important thing we can infer from figure 10 is that the
transformer takes about 22 seconds to maximise the input voltage from
10V to approximately 156V. The graph also comes out to be symmetrical
about the time axis as should be expected from a sine curve of a quantity
varying with time. Further, we can say that the curve must represent
the equation,

\begin{equation}
V_{S}=k\sin t
\end{equation}

\noindent Here, $\mathit{V_{S}}$ is the secondary output voltage,
$\mathit{t}$ is the time elapsed, and $\mathit{k}$ is the proportionality
constant, which will be a combination of the other constant parameters
listed in table 2. Note that figure 10 shows only one cycle of resonance
in the circuit. This cycle will keep repeating until the input source
is switched off.

\subsection{Voltage Ratio versus Turn Ratio}

According to equation 8, the ratio of secondary to primary voltages
of the transformers is proportional to the ratio of the number of
turns in secondary to primary coils (i.e. the turn ratio). Table 3
shows the data for voltage ratio and transformer turn ratio, and table
4 shows the values of other constant parameters. The voltage ratio
values are not exactly equal to the turn ratio. This is, again, because
the readings need to be taken manually by pausing the simulation each
time.

\begin{table}[H]

\noindent \begin{centering}
\begin{tabular}{|c|c|}
\hline 
Voltage Ratio $\left(\frac{V_{S}}{V_{P}}\right)$ & Turns Ratio $\left(\frac{n_{S}}{n_{P}}\right)$\tabularnewline
\hline 
\hline 
100  & 100.708 \tabularnewline
\hline 
200 & 200.197\tabularnewline
\hline 
300  & 300.296\tabularnewline
\hline 
400 & 400.398\tabularnewline
\hline 
500 & 500.499\tabularnewline
\hline 
600 & 600.593\tabularnewline
\hline 
700 & 700.693\tabularnewline
\hline 
800 & 800.799\tabularnewline
\hline 
900 & 901.239\tabularnewline
\hline 
1000 & 1000.998\tabularnewline
\hline 
\end{tabular}\caption{Data for voltage ratio and turn ratio.}
\par\end{centering}
\end{table}

\begin{table}[H]
\noindent \centering{}%
\begin{tabular}{|c|c|}
\hline 
Parameter & Value\tabularnewline
\hline 
\hline 
Resistance  & 22$\Omega$\tabularnewline
\hline 
Capacitance & 10$\mu F$\tabularnewline
\hline 
Primary Inductance & 4H\tabularnewline
\hline 
Input Voltage & 5V\tabularnewline
\hline 
\end{tabular}\caption{Values of constant parameters.}
\end{table}

\noindent Figure 11 below shows the plot of turn ratio versus voltage
ratio for the transformer. The curve is a straight line of the nature
$\mathit{y=mx+c}$. 

\begin{figure}[H]
\begin{centering}
\includegraphics[width=10cm,height=8cm,keepaspectratio]{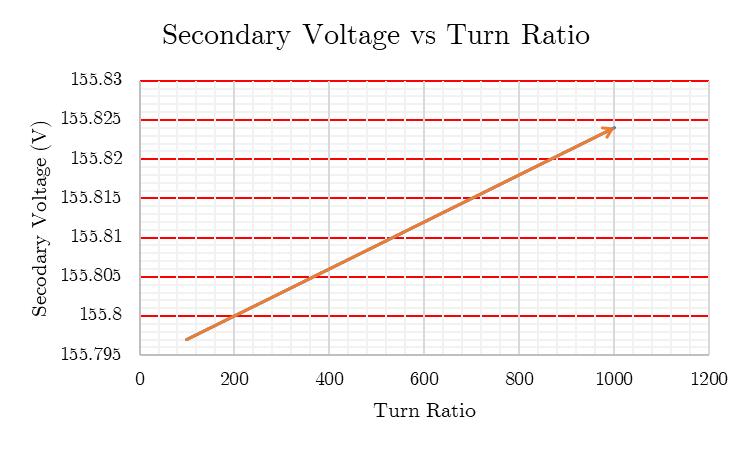}\caption{Plot of voltage ratio vs turn ratio.}
\par\end{centering}
\end{figure}

\noindent The relation between voltage ratio and turn ratio that we
get from figure 11 is shown below in equation 10.

\begin{equation}
\frac{V_{S}}{V_{P}}=0.992\frac{n_{S}}{n_{P}}\approx\frac{n_{S}}{n_{P}}
\end{equation}

\noindent The minor error may be due to the reasons stated before.
The analysis uptil this point, helps to verify already existing theory.
Now, we will shift our focus towards the main goal of finding an equation
to maximise output voltage of the slayer-exciter transformer. 

\section{Finding an Equation to Maximise SEC Output Voltage}

As aforementioned, the main purpose of the simulation is to arrive
at an equation, which can tell us (approximately, if not exactly)
how we could possibly maximize the output voltage for a given input
voltage. Apart from time ($\mathit{t}$; in $\mathit{s}$) and input
voltage ($\mathit{V_{i}}$; in V), there are a bunch of other parameters
which may or may not affect the output voltage. These are the inductance
of the primary coil ($\mathit{\mathcal{L_{\mathit{P}}}}$; in H),
parasitic capacitance ($\mathit{C}$; in $\mathit{\mu F}$), circuit
resistance ($\mathit{R}$; in $\Omega$), and the turn ratio $\left(\frac{n_{S}}{n_{P}}\right)$.
Let\textquoteright s see how the output voltage varies with each of
the above-mentioned parameters, and then we can possibly arrive at
an equation that relates the output voltage ($\mathit{V_{S}}$; in
V) to $\mathit{L_{\mathit{P}}}$, $\mathit{C}$, $\mathit{R}$ and
$\frac{n_{S}}{n_{P}}$.

\subsection{Variation of $\mathit{V_{S}}$ with $\mathit{\mathcal{L_{\mathit{P}}}}$}

From equation (7), we can say that the ratio of secondary and primary
voltages is same as the ratio of primary and secondary inductance.
Voltage is measured in volts (V) and inductance is measured in Henries
(H).

\begin{equation}
\frac{V_{S}}{V_{P}}=\frac{\mathcal{L_{\mathit{S}}}}{\mathcal{L_{\mathit{P}}}}
\end{equation}

\noindent Combining equation (11) with equation (8), 

\begin{equation}
\frac{V_{S}}{V_{P}}=\frac{n_{S}}{n_{P}}=\frac{\mathcal{L_{\mathit{S}}}}{\mathcal{L_{\mathit{P}}}}
\end{equation}

\noindent Now let us understand why increasing the primary inductance
leads to an increase in the secondary voltage even if the two quantities
are not related in any way. Suppose we increase $\mathit{\mathit{\mathcal{L_{\mathit{P}}}}}$
in equation (12). This won\textquoteright t have any effect on $\mathit{n_{S}}$,
$\mathit{n_{P}}$, or $\mathit{\mathcal{L_{\mathit{P}}}}$. But since
$\mathit{\mathcal{L_{\mathit{P}}}}$ is increased, the voltage across
the primary $\mathit{V_{P}}$ will also increase. Since $\mathit{n_{S}}$,
$\mathit{n_{P}}$ and $\mathit{\mathcal{L_{\mathit{S}}}}$ remain
constant, the only way equation (12) is satisfied, is if the secondary
voltage ($\mathit{V_{S}}$). Hence, even though the primary inductance
($\mathit{\mathcal{L_{\mathit{P}}}}$) and secondary output ($\mathit{V_{S}}$)
are not related in any way, to keep the ratio relation constant, $\mathit{V_{S}}$
increases when $\mathit{\mathcal{L_{\mathit{S}}}}$ is increased and
vice-versa. Table 5 shows the data collected for output voltage and
primary inductance and table 6 shows the values for other parameters.

\begin{table}[H]
\begin{centering}
\begin{tabular}{|c|c|}
\hline 
Secondary Voltage (V) & Primary Inductance (H)\tabularnewline
\hline 
\hline 
69.746  & 1\tabularnewline
\hline 
155.788 & 5\tabularnewline
\hline 
220.268 & 10\tabularnewline
\hline 
269.741 & 15\tabularnewline
\hline 
311.453 & 20\tabularnewline
\hline 
\end{tabular}\caption{Data for secondary voltage and primary inductance.}
\par\end{centering}
\end{table}

\begin{table}[H]
\noindent \begin{centering}
\begin{tabular}{|c|c|}
\hline 
Parameter & Value\tabularnewline
\hline 
\hline 
Resistance  & 22$\Omega$\tabularnewline
\hline 
Input Voltage & 10V\tabularnewline
\hline 
Capacitance & 10$\mu F$\tabularnewline
\hline 
Turns Ratio  & 1000\tabularnewline
\hline 
\end{tabular}\caption{Values of constant parameters.}
\par\end{centering}
\end{table}

\noindent Figure 12 below shows the plot of $\mathit{V_{S}}$ vs $\mathit{\mathcal{L_{\mathit{P}}}}$.
The primary inductance is plotted on the $\mathit{y}$-axis and the
output voltage on $\mathit{x}$-axis. The result is as expected. The
secondary voltage increases with the primary inductance.

\begin{figure}[H]
\begin{centering}
\includegraphics[width=10cm,height=8cm,keepaspectratio]{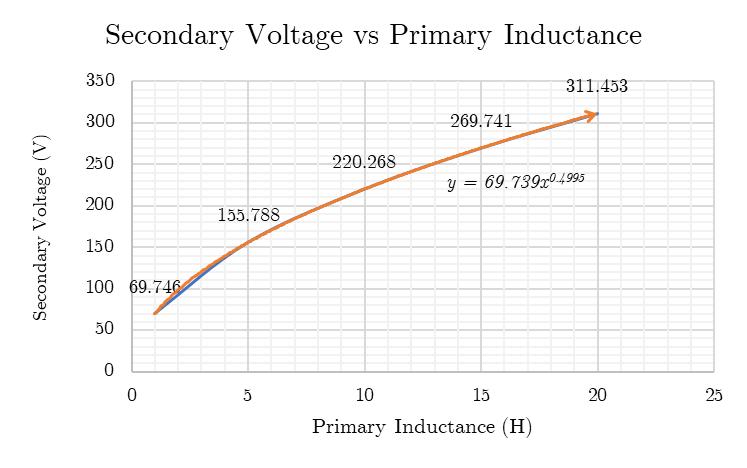}\caption{Plot for secondary voltage versus primary inductance. The orange curve
represents the most accurate trendline. The blue curve is the original
plot.}
\par\end{centering}
\end{figure}

\noindent The mathematical relation between $\mathit{V_{S}}$ and
$\mathit{\mathcal{L_{\mathit{P}}}},$ comes out as, 

\begin{equation}
V_{S}=69.739\mathcal{L_{\mathit{P}}^{\mathit{0.4995}}}\approx70\sqrt{\mathcal{L_{\mathit{P}}}}
\end{equation}

\noindent Here, $\mathit{V_{S}}$ is the output voltage at secondary
and $\mathit{\mathcal{L_{\mathit{P}}}}$ is the inductance of the
primary coil. 

\subsection{Variation of $\mathit{V_{S}}$ with $\mathit{C}$}

As discussed before, a parasitic capacitance $\mathit{C}$ (in $\mu F$)
exists naturally between the secondary coil of the transformer and
the ground. As we know, a capacitor stores electrical energy; and
for doing so, if it needs to keep the charge $\mathit{Q}$ across
it\textquoteright s plates constant, it needs to resist any increase
in voltage. In other words, capacitance is a ratio of charge and voltage.
Therefore, for a given charge, voltage decreases with increasing capacitance.

\begin{equation}
C=\frac{Q}{V_{S}}
\end{equation}

\noindent Since the parasitic capacitance is connected to the secondary
transformer, it resists any change in the secondary output voltage
($\mathit{V_{S}}$) beyond a certain maximum value. Owing to this
reasoning, we must expect that $\mathit{V_{S}}$ decreases with increasing
$\mathit{C}$. More the capacitance, more will be the resistance against
changing voltage, and hence, lesser will be the secondary voltage.
Table 7 shows the data for capacitance $\mathit{C}$ (in $\mathit{\mu F}$)
and secondary voltage $\mathit{V_{S}}$ (in V); and table 8 shows
the values of other constant parameters. 

\begin{table}[H]

\begin{centering}
\begin{tabular}{|c|c|}
\hline 
Capacitance ($\mu F$) & Secondary Voltage (V)\tabularnewline
\hline 
\hline 
1 & 279.833\tabularnewline
\hline 
2 & 213.235\tabularnewline
\hline 
3 & 179.421\tabularnewline
\hline 
4 & 157.064\tabularnewline
\hline 
5 & 140.666\tabularnewline
\hline 
6 & 129.177\tabularnewline
\hline 
7 & 119.754\tabularnewline
\hline 
8 & 111.801\tabularnewline
\hline 
9 & 105.676 \tabularnewline
\hline 
10 & 100.264\tabularnewline
\hline 
\end{tabular}\caption{Data for capacitance and secondary voltage.}
\par\end{centering}
\end{table}

\begin{table}[H]
\noindent \begin{centering}
\begin{tabular}{|c|c|}
\hline 
Parameter & Value\tabularnewline
\hline 
\hline 
Input Voltage & 5V\tabularnewline
\hline 
Resistance & 22$\Omega$\tabularnewline
\hline 
Primary Inductance & 5H\tabularnewline
\hline 
Turns Ratio & 1000\tabularnewline
\hline 
\end{tabular}\caption{Values for other constant parameters..}
\par\end{centering}
\end{table}

\noindent Figure 13 below shows the plot of variation of $\mathit{V_{S}}$
with $\mathit{C}$. The blue curve is the original plot and the orange
curve is the most accurate trend-line. Clearly, as expected, the secondary
voltage decreases with increasing capacitance.

\begin{figure}[H]

\begin{centering}
\includegraphics[width=10cm,height=8cm,keepaspectratio]{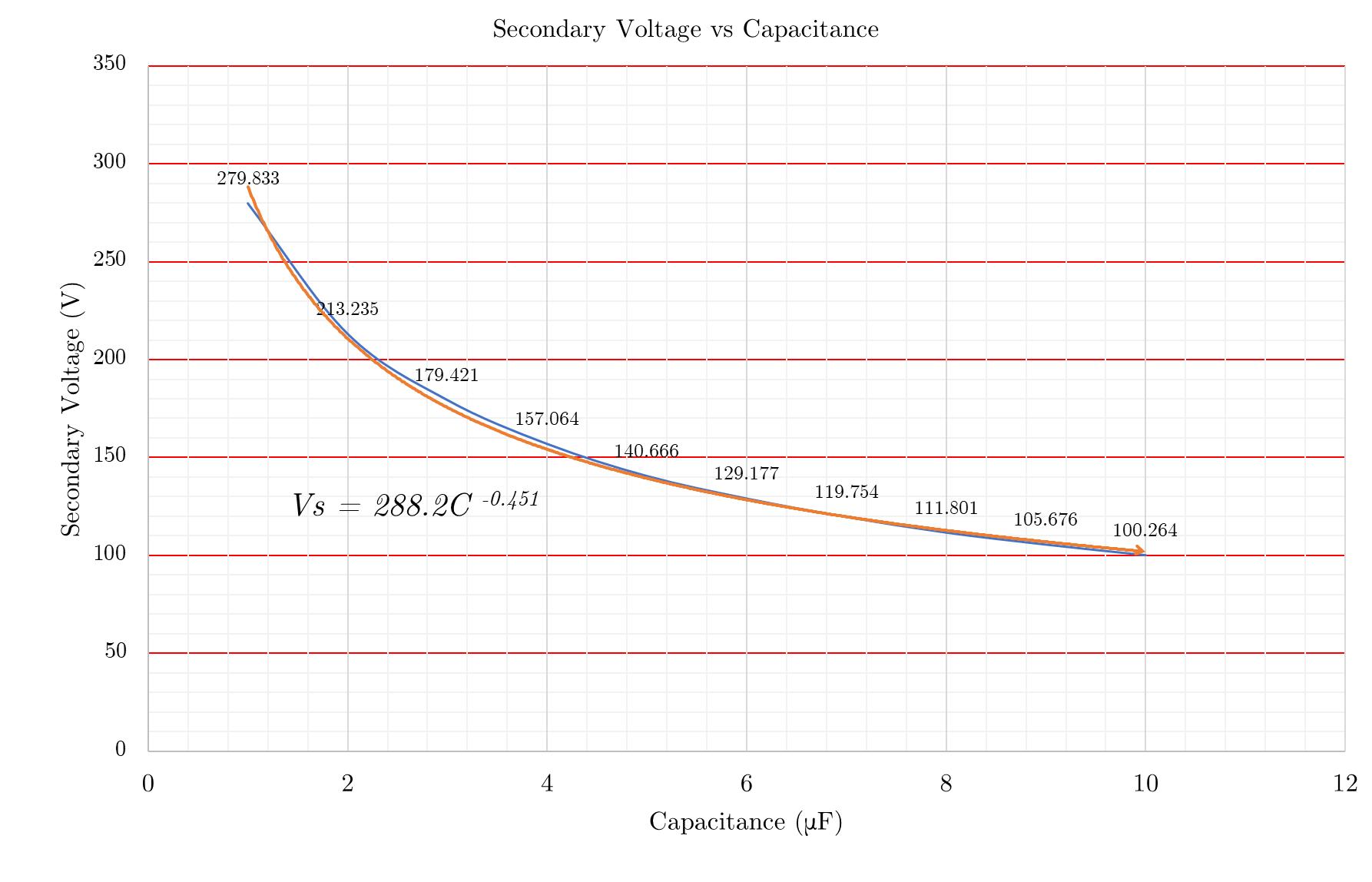}\caption{Plot of secondary voltage versus circuit resistance}
\par\end{centering}
\end{figure}

\noindent The relation between $\mathit{V_{S}}$ and $\mathit{C}$
is given from figure 13, by the equation,

\begin{equation}
V_{S}=288.2\mathit{C^{-0.451}}
\end{equation}

\noindent In equation (15), $\mathit{V_{S}}$ is the secondary voltage
and $\mathit{C}$ is the value of the parasitic capacitance. The capacitance
may again vary with the separation between the secondary coil of the
transformer and the ground. 

\subsection{Variation of $\mathit{V_{S}}$ with $\mathit{R}$}

We know that the voltage is inversely proportional to resistance.
Hence, we expect that the secondary voltage ($\mathit{V_{S}}$) will
also be related similarly to the circuit resistance ($\mathit{R}$).
Table 9 shows the data for circuit resistance (in $\Omega$) and secondary
voltage (in V); and table 10 shows the values of other constant parameters. 

\begin{table}[H]

\noindent \begin{centering}
\begin{tabular}{|c|c|}
\hline 
Secondary Voltage (V) & Resistance ($\Omega$)\tabularnewline
\hline 
\hline 
157.145  & 10\tabularnewline
\hline 
155.959 & 20\tabularnewline
\hline 
155.257 & 30\tabularnewline
\hline 
154.779 & 40\tabularnewline
\hline 
154.395 & 50\tabularnewline
\hline 
154.076 & 60\tabularnewline
\hline 
153.822  & 70\tabularnewline
\hline 
153.576  & 80\tabularnewline
\hline 
153.352  & 90\tabularnewline
\hline 
152.921  & 100\tabularnewline
\hline 
\end{tabular}\caption{Data for secondary voltage and circuit resistance.}
\par\end{centering}
\end{table}

\begin{table}[H]

\begin{centering}
\begin{tabular}{|c|c|}
\hline 
Parameter & Value\tabularnewline
\hline 
\hline 
Input Voltage & 10V\tabularnewline
\hline 
Capcitance & 10$\mu F$\tabularnewline
\hline 
Primary Inductance & 5H\tabularnewline
\hline 
Turns Ratio & 1000\tabularnewline
\hline 
\end{tabular}\caption{Values of constant parameter}
\par\end{centering}
\end{table}

\noindent Figure 14 below shows the plot for $\mathit{V_{S}}$ versus
$\mathit{R}$. The variation with resistance is not exactly of the
\textquotedblleft inverse\textquotedblright{} nature, which might
be because the resistance of the entire circuit is not taken into
consideration. Only the primary resistance provided by the resistor
has been considered.

\begin{figure}[H]

\begin{centering}
\includegraphics[width=10cm,height=8cm,keepaspectratio]{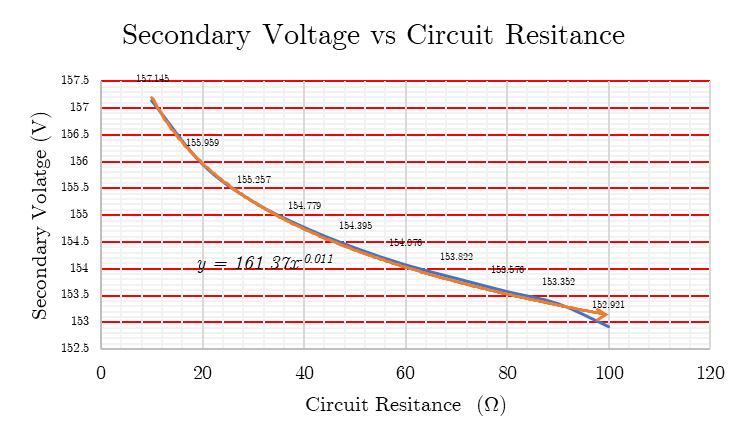}\caption{Plot of secondary voltage versus circuit resistance. The blue line
is the original plot and the orange one is the most accurate trendline.}
\par\end{centering}
\end{figure}

\noindent The relation between $\mathit{V_{S}}$ and $\mathit{R}$
is described by the sec,

\begin{equation}
V_{S}=161.37\mathit{R^{-0.011}}
\end{equation}

\noindent Here, $\mathit{V_{S}}$ is the secondary voltage and $\mathit{R}$
is the circuit resistance.

\subsection{Variation of $\mathit{V_{S}}$ with $\frac{n_{S}}{n_{P}}$}

According to equation (8), for a given primary voltage ($\mathit{V_{P}}$),
the secondary voltage ($\mathit{V_{S}}$) will increase with the turn
ratio of the transformer $\left(\frac{n_{S}}{n_{P}}\right)$, where
$\mathit{n_{S}}$ and $\mathit{n_{P}}$ are the number of turns of
the secondary and primary coils respectively. Table 11 shows the data
for turn ratio of the transformer and the corresponding voltage produced
at the secondary coil; and table 12 shows the values for other constant
parameters.

\begin{table}[H]

\noindent \begin{centering}
\begin{tabular}{|c|c|}
\hline 
Secondary Voltage (V) & Turns Ratio\tabularnewline
\hline 
\hline 
155.797  & 100\tabularnewline
\hline 
155.8  & 200\tabularnewline
\hline 
155.803 & 300\tabularnewline
\hline 
155.806  & 400\tabularnewline
\hline 
155.809 & 500\tabularnewline
\hline 
155.812  & 600\tabularnewline
\hline 
155.815 & 700\tabularnewline
\hline 
155.818 & 800\tabularnewline
\hline 
155.821 & 900\tabularnewline
\hline 
155.824 & 1000\tabularnewline
\hline 
\end{tabular}\caption{Data for secondary voltage and transformer turn ratio.}
\par\end{centering}
\end{table}

\begin{table}[H]

\begin{centering}
\begin{tabular}{|c|c|}
\hline 
Parameter & Value\tabularnewline
\hline 
\hline 
Resistance & 22$\Omega$\tabularnewline
\hline 
Capacitance & 10$\mu F$\tabularnewline
\hline 
Primary Inductance & 5H\tabularnewline
\hline 
Input Voltage & 5V\tabularnewline
\hline 
\end{tabular}\caption{Values of constant parameters.}
\par\end{centering}
\end{table}

\noindent Figure 15 below shows the plot of secondary voltage versus
turn ratio of the transformer. As expected, the relation is linear
in nature, i.e., the secondary voltage increases with the turn ratio.

\begin{figure}[H]

\begin{centering}
\includegraphics[width=10cm,height=8cm,keepaspectratio]{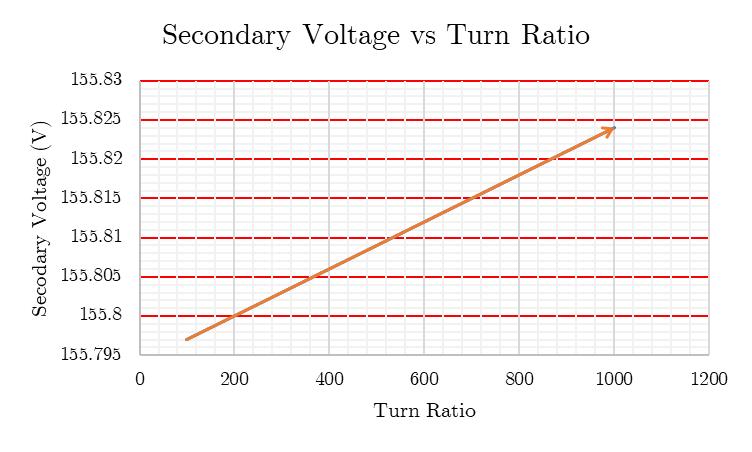}\caption{Plot of secondary voltage versus transformer turn ratio.}
\par\end{centering}
\end{figure}

\noindent The relation between the secondary voltage and turn ratio
is given from figure 15, by the equation,

\begin{equation}
V_{S}=3\times10^{-5}\left(\frac{n_{S}}{n_{P}}\right)
\end{equation}

\noindent Here, $\mathit{V_{S}}$ is the secondary voltage and $\frac{n_{S}}{n_{P}}$
is the turn ratio of the transformer.

\subsection{Equation of the SEC}

In sections 6.1. through section 6.4., we have arrived at approximate
relations between the secondary voltage ($\mathit{V_{S}}$) and other
parameters like primary inductance ($\mathit{L_{P}}$), parasitic
capacitance ($\mathit{C}$), circuit resistance ($\mathit{R}$) and
transformer ratio $\left(\frac{n_{S}}{n_{P}}\right)$. We are now
in a position to derive an equation from our experimental findings.
From equations (13), (15), (16) and (17), we see that,

\begin{align*}
V_{S} & \propto\sqrt{\mathcal{L_{\mathit{P}}}}\\
V_{S} & \propto C^{-0.451}\\
V_{S} & \propto R^{-0.011}\\
V_{S} & \propto\frac{n_{S}}{n_{P}}
\end{align*}

\noindent Combining these, we get,

\[
V_{S}\propto\frac{n_{S}}{n_{P}}R^{-0.011}C^{-0.451}\sqrt{\mathcal{L_{\mathit{P}}}}
\]

\begin{equation}
\Rightarrow V_{S}=k\frac{n_{S}}{n_{P}}R^{-0.011}C^{-0.451}\sqrt{\mathcal{L_{\mathit{P}}}}
\end{equation}

\noindent In equation (18), $\mathit{V_{S}}$ is the secondary voltage,
$\mathit{\mathcal{L_{\mathit{P}}}}$ is the primary inductance, $\mathit{R}$
is the circuit resistance and $\frac{n_{S}}{n_{P}}$ is the transformer
turn ratio. The proportionality constant ($\mathit{k}$) depends on
the accuracy of measurement. In other words, the more accurately we
measure the different parameters, the more accurate equation (18)
gets. Equation (17) is the equation we required. We call this equation
the \textquotedblleft SEC Equation\textquotedblright . One can call
it anything really, because it\textquoteright s just a name! This
equation tells us how we can maximise the output voltage at the secondary
coil of transformer, for a given source voltage, by varying different
parameters of the SEC. 

\section{The Apparent Error in the SEC Equation}

Before we get too excited by equation (18), it would be better to
see how approximate the equation is to the actual expected equation.
First of all, what is the actual expected equation? Well as we just
discussed in previous sections (equations (6), (12) and (14)), the
secondary voltage should vary with other parameters as follows:

\begin{align*}
V_{S} & \propto\mathcal{L_{\mathit{P}}}\\
V_{S} & \propto C^{-1}\\
V_{S} & \propto R^{-1}\\
V_{S} & \propto\frac{n_{S}}{n_{P}}
\end{align*}

\noindent Combining these we get, 

\[
V_{S}\propto\frac{n_{S}}{n_{P}}R^{-1}C^{-1}\mathcal{L_{\mathit{P}}}
\]

\begin{equation}
\Rightarrow V_{S}=k\frac{n_{S}}{n_{P}}R^{-1}C^{-1}\mathcal{L_{\mathit{P}}}
\end{equation}

\noindent Certainly, we can see that equation (18), which we got using
the simulation is not a good approximation of the actual equation
(19). The most immediate question that comes to mind is why does this
difference exist between equation (18) and equation (19)? Well that
might be because that measurement using the simulation is a highly
error-prone process. Measurement (as stated before) needs to be done
manually by pausing the simulation every now and then, and noting
down the various parameter values. As mentioned previously, the more
accurate the measurement, the more equatio (18) will start to look
like equation (19). 

\section{Potential Sources of Error}

There are a number of logical precautions and potential sources of
error, that may occur while preparing the physical model, as well
as the simulated model. 

\subsection{Physical Model}
\begin{itemize}
\item The connections may be improper or loose. 
\item Components may be defective. 
\item Circuit might get overheated if kept running for a long time. Due
to this, components may get damaged.
\end{itemize}

\subsection{Simulated Model}
\begin{itemize}
\item The simulation has no means to automatically export data. Hence, data
has to be obtained manually by pausing the simulation frequently.
This may result in some minor errors while measurement of parameters.
\item Components in the simulation may be improperly connected. For instance,
the Zener diode may be connected in forward bias.
\end{itemize}

\section{Results and Concluding Remarks}
\begin{itemize}
\item According to the data analysis from simulation, for a given source
voltage, the secondary voltage changes with parasitic capacitance,
circuit resistance, primary inductance and turn ratio as follows:
\[
V_{S}=k\frac{n_{S}}{n_{P}}R^{-0.011}C^{-0.451}\sqrt{\mathcal{L_{\mathit{P}}}}
\]
\item According to theory, for a given source voltage, the secondary voltage
changes with parasitic capacitance, circuit resistance, primary inductance
and turn ratio as follows:
\[
V_{S}=k\frac{n_{S}}{n_{P}}R^{-1}C^{-1}\mathcal{L_{\mathit{P}}}
\]
\item The output voltage secondary varies as a sine function of the time
elapsed.
\[
V_{S}=k\sin t
\]
\item The proportionality constant k in the above equations depends on the
accuracy of measurement of various parameters. 
\end{itemize}
\rule[0.5ex]{1\columnwidth}{0.2pt}

\end{document}